\documentstyle[12pt]{article}
\title{  Covariant Model for Relativistic\\ Three-Body Systems
\thanks{Talk given at Quark Confinement and the Hadron Spectrum VI, 
Villasimius, Cagliari, Sardinia, Italy. 21-25 September 2004}  }

\author{Philippe Droz-Vincent\\[2mm]
  LUTH, Observatoire de Paris-Meudon\\ Place Jules Janssen, F-92195 
Meudon, France}
\date{\    }

\newcommand  {\eeq}{\end{equation}}

\newcommand  {\beq}{\begin{equation} }
\newcommand  \half {  {1 \over 2} }
\newcommand  {\ytil}{\widetilde y}
\newcommand{\yhat}{\widehat y}  
\newcommand {\ztil}{\widetilde z}
\newcommand{\zhat}{\widehat z}
\newcommand  {\noi}{\noindent}
\newcommand  {\disp}{\displaystyle}
\newcommand  {\zer}{ {(0)} }

\newcommand {\alp}{\alpha} 
\newcommand {\lam}{\lambda}  
\newcommand{\Gam}{\Gamma}

\newcommand{\soulV}{\underline V}

\newcommand{\soulXi}{\underline \Xi}

\begin{document}
\maketitle
\begin{abstract}        The system is described
by three mass-shell constraints.
When at least two masses are equal, 
 this picture has a reasonable nonrelativistic limit.
At first post-Galilean order and provided the interaction is not too much 
energy-dependent, the relativistic correction is tractable like a conventional 
perturbation problem. A covariant version of harmonic oscillator is given 
as a toy model.

\end{abstract}




A system of three particles can be covariantly described by three mass-shell 
constraints, involving an interaction term referred to as {\sl potential}.
These constraints must reduce to three independent Klein-Gordon (or Dirac) 
equations in the absence of potential. In any case, they determine the
 evolution of a wave function which depends on three four-dimensional
 arguments, say $p_a$ with  $a,b  = 1,2,3, $
 if we chose the momentum representation of quantum mechanics. 


\noi
Naturally, the potential depends on both configuration and momentum variables, 
$q_a , p_b$,  and must allow for mutual compatibility of the constraints.
Moreover it happens that, just like in the Bethe-Salpeter approach, manifest 
covariance is paid by the presence of redundant degrees of freedom of which
 the elimination is by no means straightforward (in contrast to the 
two-body case).
These two important issues have been considered earlier by 
H.~Sazdjian~\cite{saz} who aimed at solving the general $n-$body case and 
proposed an approximate solution.

\noi
Specially dealing with the {\sl three-boson} case, we have recently exhibited  in
 closed analytic form a new set of variables $q' _a , p' _b$.
In terms of these new variables, admissible expressions for the  potential are 
 explicitly available, 
 and two superfluous degrees of freedom can be eliminated \cite{droz}.
Setting  $ P = \sum p$  
we linearly introduce relative variables
$$   z_A = q_1 - q_A ,  \qquad   y_A =  {P \over 3 } - p_A , 
\quad     \qquad    A =  2,3           $$
and similar formulas for  $z'_A , y'_B $ in terms of $q'_a , p' _b$.

\noi
The mass-shell constraints can be equivalently replaced by their sum and 
differences; it is convenient to set
$\disp    \nu_A  = \half (m_1 ^2  -  m _ A ^2 ) $.

The difference 
equations, in their original form, yield no simplification. But we perform a 
{\em quadratic} change among the momenta,  say  $p_a  \rightarrow p'_a$, 
or equivalently           $P, y_A \rightarrow  P' , y' _A$. 
in order to ensure the elimination of two redundant  degrees 
of freedom; this change is characterized by   
$$ (p_1 -  p_A ) (p_1 + p_A )  =   (p'_1 -p' _A ) \cdot P    $$
whereas $P'=P$  and the transverse parts of the momenta  remain unaffected,
 say ${\ytil}'  =  \ytil$, where the {\sl tilda} on any four-vector refers to its
   transverse part  with respect to $P$.
  
\noi Of course, this procedure generates 
a change of canonical variables~\cite{droz},
 in particular we  obtain new configuration variables,   $z' _A$.
\medskip
  
{\sl Three-dimensional Reduction}

\noi   We impose a sharp value of the  total linear momentum, it is a timelike 
vector $k$, and we define  $k \cdot k = M^2$.

\noi {\sl Notations:} 
The {\em hat} on any vector refers to its transverse part  with respect to $k$.

\noi {\em Underlining} any  dynamical variable indicates that, in its expression, 
 we substitute $k$ for $P$  and take into account equation 
the difference equations
\beq 3  y' _A   \cdot k \   \Psi =  (4 \nu _A - 2 \nu _B ) c^2 \   \Psi
                                         \label{diffeq}     \eeq
We  factorize out the relative energies; as a result the  sum equation becomes
  \beq   (3 \sum m^2 - M^2) c^2  \        \psi =
 6 (\yhat _2 ^2  + \yhat _3 ^2  +  \yhat _2 \cdot  \yhat _3 ) \psi
+  (6  M^2 c^2 \soulXi    +  18  \soulV )  \     \psi  
                       \label{sumeq}     \eeq
for a {\em reduced} wave function  $\psi$ which depends only on the transverse 
relative momenta   $ \disp   \yhat '_A      =  \yhat_A $.

\noi 
The meaning of  $\Xi $  is purely kinematic; this term  depends only on the momenta and 
can be expressed in terms of their transverse part and $P$.
Here  $V$  denotes  the relativistic potential; it may be phenomenological or 
motivated by considerations of field theory. In particular it may be {\em formally} 
 constructed     as    a sum of two-body terms, like in equation  (\ref{oscar}) below;
so doing one uses {\em the shape} of two-body potentials but (for the sake of compatibility)
 with the {\em new} three-body variables as arguments.
  Not only the total momentum $P$ but also the  new configuration variables
  $z' _A$ mix the two-body  clusters, which  amounts to automatically incorporate 
  three-body forces.
           Admissible potentials entail
 that $\soulV$ is a  function of the  {\em new} variables 
  $\zhat ' _2 , \zhat ' _3$ and $M^2 c^2$.
                      
 \noi 

\noi     The  reduced equation  (\ref{sumeq}) is actually 
 a nonconventional   eigenvalue probem, where the operator to be 
diagonalized {\em explicitly  depends} on its eigenvalue. This situation is by no 
means a special drawback of our model, in fact it is common in relativistic
 quantum mechanics~\cite{rizov}, but it would make a  general treatment rather 
involved.  

\noi On the other hand, it is natural to  expand the formulas in powers of 
$1/c^2$ and to look for the nonrelativistic limit.  For arbitrary masses, 
the term     $ M^2 c^2 \soulXi$ generally  blows up,  which  leads to consider, 
instead of (\ref{sumeq})  an alternative combination of the mass-shell constraints.

\medskip

{\sl Two equal masses}.

\noi   Drastic simplifications arise when two masses are equal, say
$ m_2 = m_3 =m$, equivalently  $\nu_2 = \nu _3 = \nu$.
We find  that
the Galilean limit of our eigenvalue problem is  a  Schroedinger equation
 with effective (or {\em Galilean}) masses that are generally distinct from 
the constituent masses $m_a$.
However  they still  coincide with the constituent masses,  at first order
  in the  "mass-dispersion index"   $\nu /  m^2$.

\medskip

{\sl Three Equal Masses}.

\noi When $ m_a =m$ for all particles, equation (\ref{sumeq})  can be written
as follows, using the rest frame
\beq   \lam  \psi =  ({\bf y}_2 ^2 +  {\bf y}_3 ^2 
  +  {\bf y}_2  \cdot   {\bf y}_3  )    \psi
- 3  \soulV   \psi   -  M^2   c^2  \soulXi   \psi    \label{sumeq3m}     \eeq
with  $   6   \lam  =    (M^2 - 9  m^2)  c^2   $.
Now the last term in  (\ref{sumeq3m}) 
remains finite in the nonrelativistic  limit. Indeed we can write 
$\disp  M c^2  \soulXi  =   {1 \over M^2 c^2}      \Gam _{(0)}
  +  O(1/c^4 )    $  where   
 \beq     \Gam _{(0)} =  {3\over 4}   \      \{ 
({\yhat _2}^2 )^2 +  ({\yhat _3}^2 )^2 + 4   (\yhat _2 \cdot \yhat _3 )^2
+ 2  ({ \yhat _2}^2  +  { \yhat _3}^2 ) \      ( \yhat _ 2 \cdot \yhat _3)
 - {\yhat _2 }^2  {\yhat _ 3 }^2    \}     
                           \label{Gamzero}                    \eeq   
At first order in $1/c^2 $ we can, in $\soulXi$, replace $M^2$ by $9m^2$, 
which is independent from $\lam$.
Thus we replace $M^2 c^2 \soulXi$ by    $ \Gam _{(0)}  /  9m^2 c^2$. 
If the relativistic "potential" $V$  doesnot depend on $P^2$, or if this 
dependence is of higher order,  equation  (\ref{sumeq3m})
  becomes a conventional eigenvalue problem, tractable by 
perturbation theory.  The last term in(\ref{sumeq3m}) brings a negative 
correction to the value   $\lam_{\rm NR}$ furnished by the nonrelativistic 
approximation, say 
$$ \lam = \lam_{NR} - < \Gam _{(0)} >   / 9 m^2 c^2           $$
if $\lam _{\rm NR}$ corresponds to  a nondegenerate level.
One has to calculate 
$ < \Gam _\zer > $ in the eigenstate solution of the nonrelativistic problem.

\medskip

{\sl  Harmonic Oscillator}

\noi A covariant version of the harmonic potential is given by
\beq  V = 2  \kappa   
\     \{  ({\ztil}'_2  ) ^2  +   ( {\ztil}'_3  )^2  -   
       {\ztil}'_2   \cdot    {\ztil}'_3      \}       \label{oscar}    \eeq
hence $\soulV$ in terms of
 $\zhat '_A \cdot \zhat ' _B =   -   {{\bf z}'_A} ^2  \cdot  {{\bf z}'_B} ^2 $.
In the nonrelativistic limit we recover the naive $SU_6$ invariant 
Schroedinger equation.  At the first post-Galilean approximation, 
$M^2 $ can be replaced by   $9m^2$,  neglecting  the dependence on 
total energy in the reduced equation. At this stage, the eigenvalue problem amounts to 
diagonalize a nonrelativistic harmonic oscillator, with potential
$\disp    V_{\rm NR} =  - 3 \soulV  /  m $,  submitted to a
 momentum-dependent perturbation. 
Expressed in terms of  Jacobi-like coordinates, namely   
$\disp   {\bf R  }_2    = -   {\bf z}' _2   +   {\bf z}' _3           \qquad  \quad
  {\bf R}_3   =    
( {\bf z}' _2   +   {\bf z}' _3  )  /     \sqrt{3}  $
and their conjugate momenta,  the unperturbed  ground state is a Gaussian.
If the unit of lenght is choosen  such that
$\disp      \kappa  =  {2 \over 9}$,
 one finds
     $ \disp    < \Gam _{(0)} > =  11 + 1/4     $.

\medskip
 
 \noi  This approach is intented for applications to confining 
interactions; future  work should implement spin and  investigate a 
possible contact with     
recents developments \cite{bij} of the BS approach.

                                                           


\vskip 0.3cm


\begin{thebibliography}{19}
\providecommand{\bibinfo}[2]{#2}




\bibitem{saz}
H. Sazdjian,
\newblock{\em Physics Lett.} {\bf B 208}, 470
 (1988); 
\newblock
{Annals of Phys.}       {\bf 191}, 52
 (1989).

\bibitem{droz}
Ph. Droz-Vincent,
\newblock {\em Int. Jour. of Theor. Phys.}
 {\bf 42}, 1809 (2003).


\bibitem{rizov}
V.A. Rizov, H. Sazdjian, I.T.Todorov
\newblock {\em Ann. of Phys.} {\bf 165}, 
59,  (1985)

\bibitem{bij}
J. Bijtebier, 
\newblock{\em Jour. of Phys. G: Nucl. Part. Phys.}
{\bf 26}, 871
(2000) 



       


\end{thebibliography}
\end{document}